\documentclass[journal]{IEEEtran}

\usepackage{upgreek}

\ifCLASSINFOpdf
   \usepackage[pdftex]{graphicx}
   
\else
\fi
\begin{document}
%
\title{Microresonator frequency reference for \\ terahertz precision sensing and metrology}
%
%
%

\author{Rishabh Gandhi,~
        Rainer Leonhardt,~
        Dominik~Walter~Vogt

\thanks{Rishabh Gandhi, Rainer Leonhardt, and Dominik Walter Vogt are with the Department of Physics, The University of Auckland, Auckland 1010, New Zealand, and the Dodd-Walls Centre for Photonic and Quantum Technologies, New Zealand (e-mail:d.vogt@auckland.ac.nz)}}
\maketitle

\begin{abstract}
Highly sensitive terahertz (THz) sensors for a myriad of applications are rapidly evolving. A widespread sensor concept is based on the detection of minute resonance frequency shifts due to a targeted specimen in the sensors environment. Therefore, cutting-edge high resolution continuous wave (CW) THz spectrometers provide very powerful tools to investigate the sensors' performances. However, unpredictable yet non negligible frequency drifts common to state-of-the-art CW THz spectrometers limit the sensors' accuracy for ultra-high precision sensing and metrology. Here, we overcome this deficiency by introducing an ultra-high quality (Q) THz microresonator frequency reference. Measuring the sensor's frequency shift relative to a well-defined frequency reference eliminates the unwanted frequency drift, and fully exploits the capabilities of modern CW THz spectrometers as well as THz sensors. In a proof-of-concept experiment, we demonstrate the accurate and repeated detection of minute resonance frequency shifts of less than 5\,MHz at 0.6\,THz of a THz microresonator sensor.     
\end{abstract}

\begin{IEEEkeywords}
Terahertz, microresonator, continuous wave spectroscopy, frequency reference
\end{IEEEkeywords}

%
\IEEEpeerreviewmaketitle

\section{Introduction}
%
%
%
%
\IEEEPARstart{T}{erahertz} sensors for sophisticated applications in microfluidics, gas-phase spectroscopy, thin-film sensing, and bio-molecular sensing to name but a few are rapidly evolving \cite{cao2020tunable,lee2017nano,s20103005,serita2019terahertz, xu2017mechanisms,banerjee2019ultra}. A common modality of these THz sensors relies on an induced resonance frequency shift upon a physical change in the system. Ultimately, the smallest detectable frequency shifts are limited by the linewidth of the resonance feature. With continuous improvement in THz sensors, their resonance features are becoming increasingly narrower, allowing to sense minute changes in the system. In particular, rapid advancements in the field of THz microresonator sensors recently culminated in the development of a sub-wavelength thin ultra-high-Q THz disc microresonator. The disc microresonator shows an unprecedented Q-factor of more than 120,000 at 0.6\,THz, corresponding to an intrinsic linewidth $<$\,5\,MHz, alluding to the possibility to resolve minute resonance frequency shifts of a few megahertz \cite{thin_disc_paper,s20103005}.

However, to fully unleash the capabilities of THz sensors with ultra narrow resonance features, a sufficiently stable THz spectrometer with a very high frequency resolution is imperative. While state-of-the-art CW THz spectrometers provide frequency resolutions of up to 1\,MHz \cite{deninger20152}, they experience an unpredictable yet non-negligible frequency drift over time under common environmental conditions. This frequency drift compromises the accuracy of measurements of induced resonance frequency shifts, limiting the breakthrough of highly sensitive THz sensors. 

\begin{figure*}[tb]
\centering
\includegraphics[width=2\columnwidth]{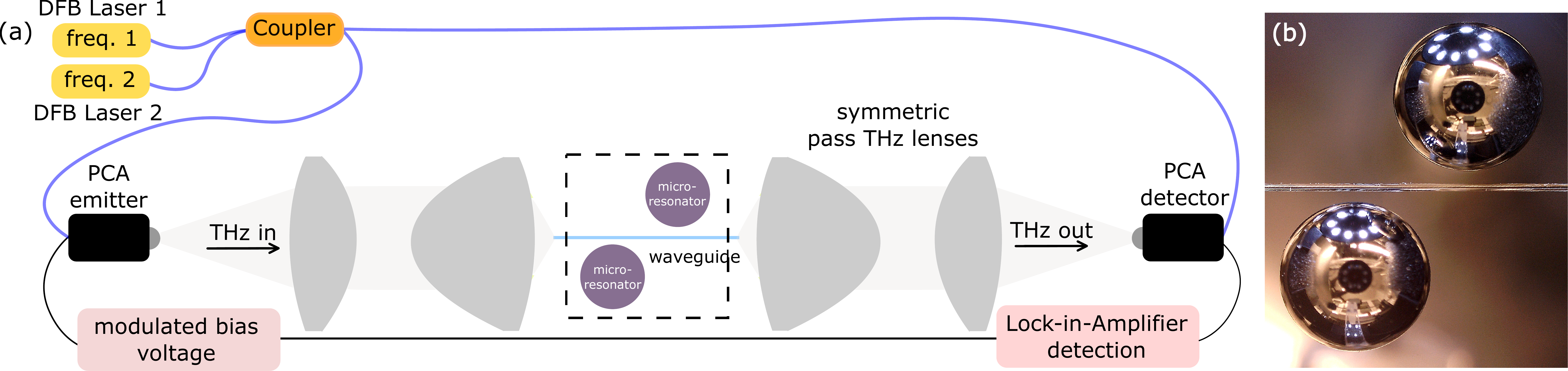}
\caption{(a) Schematic of the experimental setup. The positions of the HRFZ-Si THz microresonators are controlled with 3D translation stages. The THz microresonator sensor is tilted relative to the air-silica step-index waveguide using a goniometer. The setup is in a \,4\,$\%$ relative humidity atmosphere to minimize distortions from water vapour. (b) The microscope image shows the two 4\,mm diameter HRFZ-Si spherical microresonators and the 200\,$\upmu$m diameter waveguide.}
\label{fig:1}
\end{figure*}

Here, we will overcome this deficiency by leveraging the ultra-narrow resonance features of THz microresonators to develop a novel frequency reference for THz spectroscopy. The ultra-high-Q THz microresonators are ideal as a frequency reference due to their inherently well-defined resonance frequencies \cite{vogt2018ultra}. Measuring an induced resonance frequency shift of an ultra-high Q THz sensor relative to a well-defined frequency reference eliminates the uncertainty of an unknown frequency drift inherent to modern CW THz spectrometers, significantly improving the sensitivity of cutting-edge THz sensors. This novel high-precision THz frequency reference will not only significantly propel the development of a new suite of THz sensors but also benefits CW THz spectroscopy in general. 





\section{Methods}

We demonstrate the frequency reference's capabilities by measuring minute resonance frequency shifts of a spherical ultra-high Q THz microresonator sensor. In particular, we investigate the resonance frequency shifts that are induced by tilting the nearly perfectly spherical THz microresonator, while maintaining the plane of excitation: the microresonator's intrinsic resonance frequencies are directly proportional to the effective diameter experienced by the microresonator's modes, therefore detecting minute spherical imperfections. In addition to representing a convincing demonstration of the frequency reference's capabilities, the presented results pave the way for unprecedented THz precision metrology.   

Both the THz microresonator sensor and the frequency reference are spherical microresonators made of high-resistivity float-zone grown silicon (HRFZ-Si) with a resistivity $>$\,10k$\Omega$cm. The spherical HRFZ-Si microresonators are extremely robust, easy to handle and provide intrinsic Q-factors in excess of 60,000, corresponding to an intrinsic linewidth of about 10\,MHz \cite{vogt2018ultra}. The choice of the sphere's diameter determines the resonance frequencies and free-spectral range (FSR) of the microresonator. Without loss of generality, we are using 4\,mm diameter spheres for both the THz microresonator sensor and the frequency reference, with a less than 1\,$\%$ difference in diameter. The FSR of both THz microresonators is about 9\,GHz \cite{vogt2018ultra}.

A schematic of the experimental setup, including a microscope image of both THz microresonators, is shown in Fig. \ref{fig:1} (a) and (b), respectively. Coherent terahertz radiation is generated and detected using a CW THz spectrometer with fiber-coupled photo-conductive antennas (TeraScan 1550) \cite{deninger20152}. The spectrometer provides a maximal frequency resolution of 1\,MHz. Free-space THz radiation is collimated and focused using specially developed symmetric-pass polymer lenses \cite{lo2008aspheric}. A 200\,$\upmu$m diameter single-mode air-silica step-index waveguide is used for evanescent coupling to the THz microresonators. The microresonators are mounted on 3D precision translation stages to control the positions (i.e. the coupling strengths) of the microesonators relative to the waveguide. Additionally, the THz microresonator sensor is mounted on a goniometer, allowing to tilt the sphere of up to $\pm$\,6$^{\circ}$ from the vertical axis. The THz signal transmitted from the waveguide is analysed using a Hilbert Transform. This techniques provides a very powerful data analysis for CW THz spectroscopy, allowing to retrieve the analytic signal (amplitude and phase information) of the detected THz photocurrent at every measured frequency step \cite{vogt2019coherent,vogt2017high}. Finally, the entire setup is placed in an enclosure with a controlled relative humidity of 4\,$\%$ in order to minimise distortions from water vapour.             

The measurement procedure is as follows: First, we measure the THz signal in the frequency range of interest, without the frequency reference and THz sensor (reference scan). Comparing all subsequent scans to a reference scan allows to eliminate any contributions from unwanted features in the measurements like standing waves in the THz path \cite{vogt2019coherent}. Next, we position the frequency reference sufficiently close to the waveguide to achieve close to critical coupling (frequency reference scan) \cite{vogt2017terahertz}: close to critical coupling, the waveguide transmission approaches zero, providing the largest signal to noise ratio. After the initial adjustment, the frequency reference's position stays constant during the remaining measurement procedure. Analysing the intensity ratio and phase difference of the reference scan and frequency reference scans allows to establish the resonance frequency of the loaded frequency reference. This resonance frequency is the frequency reference for all subsequently measured resonance frequency shifts.

If the THz microresonator frequency reference is in a stable environment, its intrinsic reference frequency is fixed, and any change in apparent resonance frequency is due to a frequency drift of the CW THz spectrometer. Of course, changes in the environment of the THz microresonator frequency reference, like coupling position, temperature and humidity, can change the microresonators resonance frequencies. Consequently, great care needs to be taken to stabilise and monitor the resonators environment. However, when the analytic dependence of a specific environmental parameter on the resonance frequency is known, it can be compensated for in the data analysis. For example, with the linear thermal drift rate of the HRFZ-Si THz microresonators well known (27\,MHz/K at 0.6\,THz) \cite{vogt2018thermal}, the resonance frequency drift upon a known resonator temperature change can readily be compensated for. Also, if the frequency reference and the THz sensor are identical, any change in the environment that effects both microresonators equally, will cancel and not contribute to a change in the frequency difference between the sensor and frequency reference.   


\begin{figure*}[htb]
\centering
\includegraphics[width=2\columnwidth]{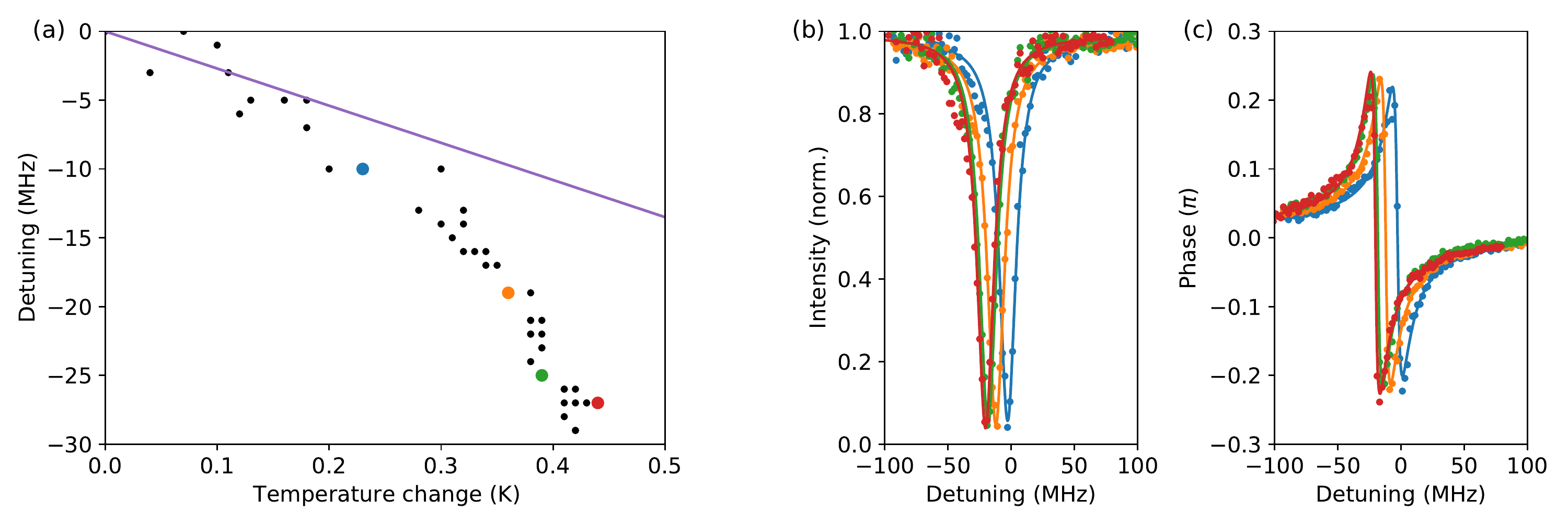}
\caption{(a) Frequency drift of the frequency reference (dots) as a function of resonator temperature. The purple line shows the expected thermal drift of the resonance frequency \cite{vogt2018thermal}. (b) and (c) show exemplary the intensity and phase information (dots: measurement; solid lines: fitted analytical model) for the blue, orange, green and red dots in (a).}
\label{fig:2}
\end{figure*}

Once the frequency reference's resonance frequency is established, the sensing measurements can be performed. Here, we first perform a scan with the weakly coupled THz microresonator sensor and the close to critically coupled frequency reference (sample scan). Analysing the intensity ratio and phase difference of the reference scan and sample scan allows to extract the frequency difference between the THz sensor (for this particular tilt and coupling position) and the frequency reference. Because the loaded resonance frequency of the THz microresonator sensor shows a small dependence on the coupling position ($<$5\,MHz) \cite{vogt2019anomalous}, and to improve the accuracy of the measurements, we extract the resonance frequencies for several coupling positions at various distances to the waveguide. In particular, the resonance frequency shift follows an exponential behaviour as a function of the distance between the micoresonator and waveguide \cite{gorodetsky1999optical}. Fitting the experimental data with an exponential function allows to determine the intrinsic resonance frequency of the THz microresonator sensor (i.e. at very large distances between the microresonator and the waveguide implying very weak coupling). Next, we repeat this process, but for a different tilt of the THz microresonator sensor. A change in frequency difference between the intrinsic resonance frequencies at different tilts compared to the frequency reference will unequivocally show small imperfections of the HRFZ-Si sphere. Finally, we repeat the measurements for the initial tilt of the THz microresonator sensor to establish the reproducibility and accuracy of the proposed frequency reference technique for THz spectroscopy.

\section{Results and Discussion}

The THz microresonator frequency reference provides an inherently well-defined resonance frequency, serving as a precise reference for induced frequency shifts of THz sensors. In particular, the frequency reference eliminates the unwanted frequency drift of state-of-the-art CW THz spectrometers as exemplary demonstrated in Fig. \ref{fig:2}. Fig. \ref{fig:2} (a) shows the frequency drift (black dots) of the frequency reference resonance frequency at 0.6035\,THz, while monitoring the air temperature in the resonator's environment over a period of about seven hours. The CW THz spectrometer was running continuously for four days prior to the shown measurements, to ensure sufficient warm-up time for the spectrometer's laser heads and electronics. The intensity and phase measurements (dots) at the blue, orange, green and red dots in Fig. \ref{fig:2} (a) are exemplary shown in Fig. \ref{fig:2} (b) and (c), respectively; the corresponding fits with the complex analytical model are shown with solid lines. Please note, because the setup is in an enclosure, and the temperature change is very slow (over several hours) it is reasonable to assume, that the air temperature and resonator temperature are in equilibrium \cite{vogt2018thermal}. Also, the temperature increase inside the enclosure is caused by a room temperature increase. During the measurement time of about seven hours, the temperature of the resonator changes by about 0.4\,K and the resonance frequency red-shifts by about 30\,MHz. While a frequency red-shift with increase in environmental temperature is expected [27MHz/K \cite{vogt2018thermal}, purple solid line in Fig. \ref{fig:2} (a)], the recorded frequency drift strongly deviates from the expected behaviour. Since no other environmental parameters changed (the humidity in the setup enclosure is closely monitored), the observed frequency drift on top of the expected thermal drift must originate from the CW THz spectrometer. This CW THz spectrometer specific frequency drift is common to both the frequency reference and the THz sensor, and because the sensor's frequency shift is measured relative to the frequency reference, does not deteriorate the accuracy of the measurements. While a frequency drift of the CW THz spectrometer of about 20\,MHz (on top of the microresonator's thermal drift) is negligible for many THz applications, it is considerable for high precision sensing and metrology applications as demonstrated below. Moreover, here, because the frequency reference and THz sensor are made from the same material, the observed thermal drift equally effects both resonators and does not impact the measurement results. However in general, for a different THz sensor - because the thermal drift of the HRFZ-Si THz microresonator frequency reference is well known - it can readily be accounted for in the subsequent data analysis. Also, Fig. \ref{fig:2} clearly establishes a correlation between the CW THz spectrometer's frequency drift and the room temperature, indicating that a precisely controlled environmental temperature of the entire spectrometer could potentially minimise the observed frequency drift. However, while control of the spectrometer's temperature better than 0.1\,K is technically very challenging, it is easily addressed with the THz microresonator as frequency reference.



\begin{figure*}[htb]
\centering
\includegraphics[width=2\columnwidth]{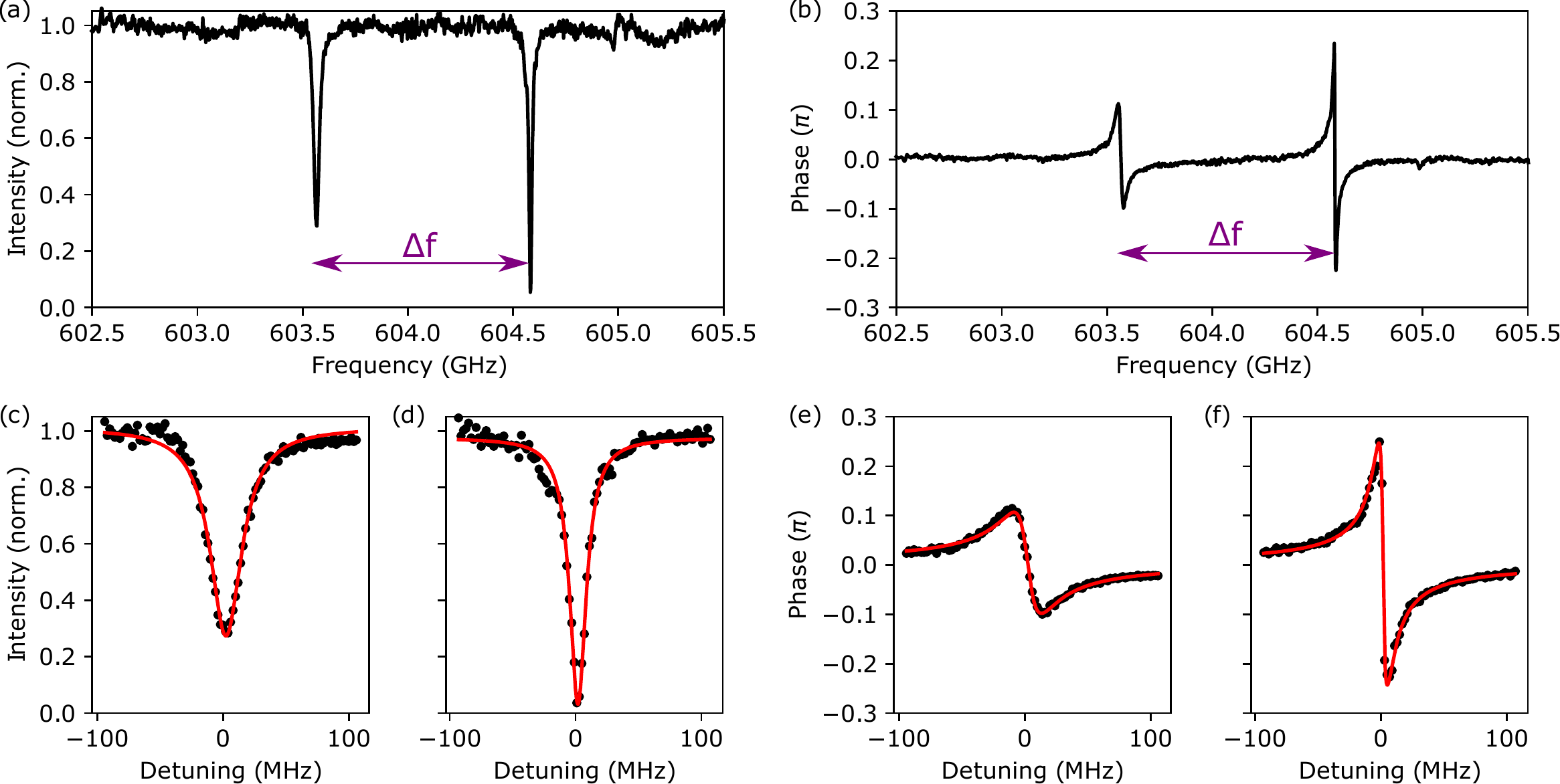}
\caption{(a) Intensity and (b) phase measurements of the sample and frequency reference resonances at 0.6035\,THz and 0.6045\,THz, respectively. (c) to (f) show the corresponding zoom-ins, with the fitted analytical model shown with red solid-lines.}
\label{fig:3}
\end{figure*}


Figure \ref{fig:3} shows exemplary a frequency scan from 0.6025\,THz to 0.6055\,THz of the measured intensity (a) and phase (b) of both the THz microresonator sensor at 0.6035\,THz and the frequency reference at 0.6045\,THz for one particular coupling position and tilt. The frequency difference $\Delta f$ is about 1\,GHz between the sensor and the frequency reference as indicated with the purple arrow in Fig. \ref{fig:3} (a). The frequency reference is close to critical coupling, while the THz microresonator sensor is only weakly coupled \cite{vogt2018prism}. Please note, that there is no particular reason why we have chosen those two resonance for this experiment. Of course, the frequency reference has many resonance frequencies allowing to choose the most suitable, providing a significant flexibility for a range of applications. Figures \ref{fig:3} (c) and (d) show zoom-ins of the intensity and phase, respectively, for both resonances. The fitted analytical model (solid lines), used to extract the resonance frequencies, shows a very good agreement with the measurements. 


\begin{figure}[b!]
\centering
\includegraphics[width=\columnwidth]{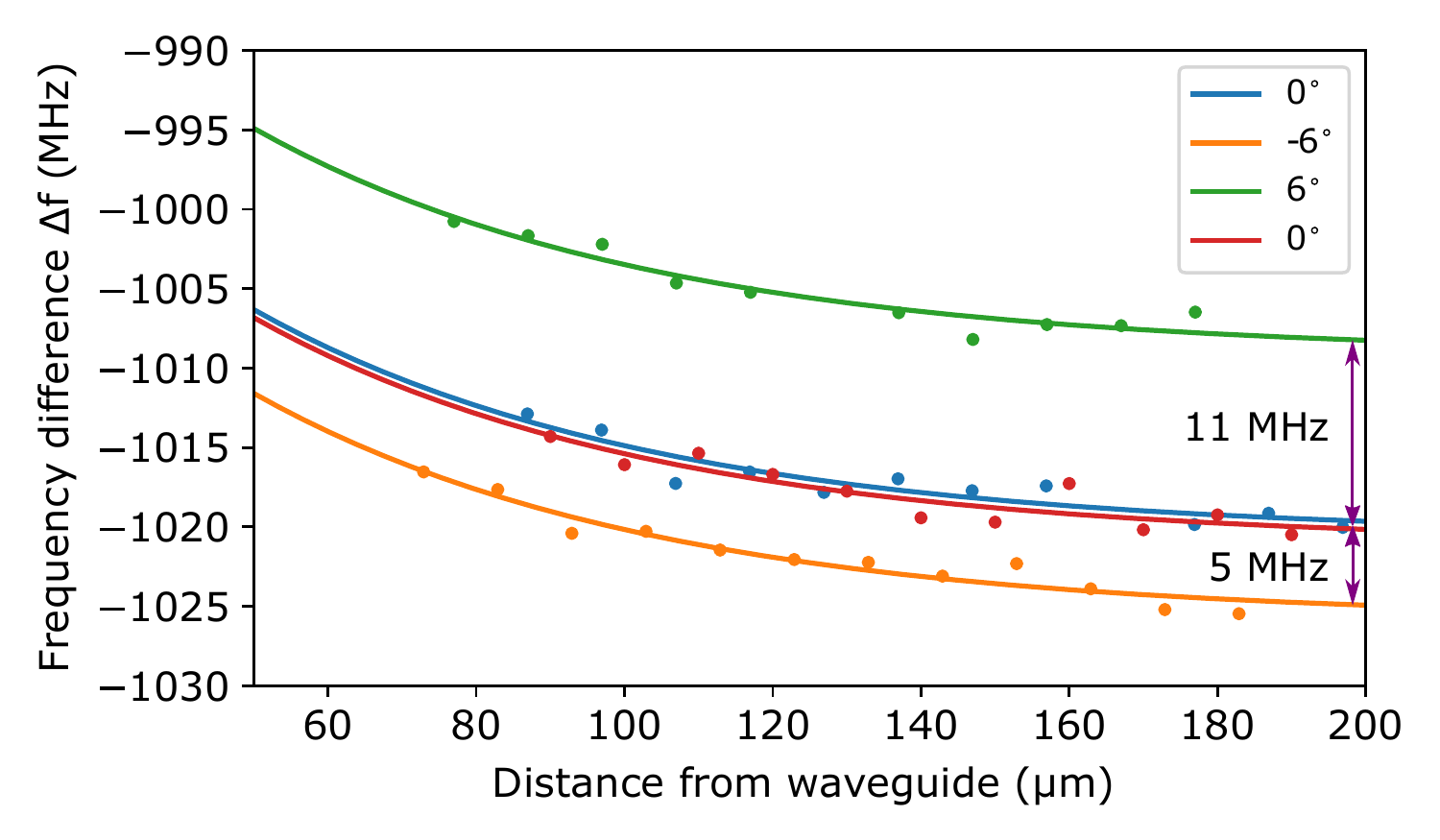}
\caption{Measured frequency difference (dots) between the frequency reference and the THz microresonator sensor for four tilts (0$^{\circ}$,-6$^{\circ}$, 6$^{\circ}$, 0$^{\circ}$), and the corresponding exponential fits (solid lines). The purple arrows indicate the frequency difference of about 11\,MHz for the 6$^{\circ}$ tilt and 5\,MHz for the -6$^{\circ}$ compared to the 0$^{\circ}$ tilt; these values are calculated for an infinite waveguide to resonator distance.}
\label{fig:4}
\end{figure}


As described above, we repeatedly extract $\Delta f$ for each tilt at different coupling positions, and extract the intrinsic frequency difference from the exponential fit. Figure \ref{fig:4} shows the measurements and corresponding fits for four tilts [0$^{\circ}$(shown in blue), -6$^{\circ}$(orange), 6$^{\circ}$(green), 0$^{\circ}$(red)]. For each tilt, we performed 10 measurements at increments of 10\,$\upmu$m coupling distance, and the exponential fits show excellent agreement with the measurements. Please note, that the exponential fits are constrained to the same decay rate for all data sets, as the resonant mode with the same evanescent field is excited for all tilts of the spherical microresonator. Also, the distance from the waveguide is normalised to the first data set for the 0$^{\circ}$ tilt to facilitate an easier comparison between tilts, as the absolute distance from the waveguide is irrelevant for the determination of the intrinsic resonance differences. From the measurements, it can be clearly seen, that both the -6$^{\circ}$ tilt and 6$^{\circ}$ tilt show a shifted frequency difference compared to the 0$^{\circ}$ tilt. In particular, the -6$^{\circ}$ tilt and the 6$^{\circ}$ tilt, show an intrinsic frequency difference of 5\,MHz $\pm$ 1\,MHz and 11\,MHz $\pm$ 1\,MHz, respectively, compared to the 0$^{\circ}$ tilt. Most importantly, repeated measurements at the start (blue) and end (red) of the measurement process of the 0$^{\circ}$ tilt agree within 0.5\,MHz, unequivocally demonstrating the impressive accuracy that can be achieved with the frequency reference. Due to the frequency drift of the CW THz spectrometer of about 20\,MHz (see Fig. \ref{fig:2}), it would have been impossible to resolve these minute frequency shifts without the frequency reference. In particular, with a time difference of about seven hours between the first and last measurement for the 0$^{\circ}$ tilt, the presented results clearly highlight the long-term stability and accuracy of the measurements using a THz microresonator frequency reference.           

Comparison with finite-element simulations (COMSOL Multiphysics\textsuperscript{\textregistered} software \cite{COMSOL}) reveals that the frequency shifts of 5\,MHz and 11\,MHz correspond to a change in effective diameter of the HRFZ-Si sphere of 18\,nm and 40\,nm, respectively. 18\,nm is about 28,000 times smaller than the free-space wavelength at 0.6\,THz, highlighting the capabilities of the THz microresonator sensor combined with the frequency reference to detect minute imperfections in the HRFZ-Si sphere that are significantly smaller than the free-space wavelength.   

\section{Conclusion}
We have shown that for ultra precise THz sensing and metrology applications a well defined frequency reference is essential as current state-of-the-art CW THz spectrometers are not stable enough for measurements that require megahertz accuracy. In particular, the presented results unequivocally demonstrate the capabilities of an ultra-high Q THz microresonator as a frequency reference, allowing to detect frequency shifts in the order of 3\,MHz in the THz frequency generated by the CW THz spectrometer. For example, minute variations in the effective diameter (less than ${10}^{-5}$) of a nearly perfect sphere can be accurately detected due to the well determined resonances of the frequency reference. Ultimately, the frequency reference overcomes the unwanted frequency drift common to modern CW THz spectrometers, paving the way for high precision THz sensing and metrology applications.       

\ifCLASSOPTIONcaptionsoff
  \newpage
\fi



\bibliographystyle{IEEEtran}
\bibliography{ref}
\end{document}